\documentclass[aps,showpacs,twocolumn]{revtex4}
\usepackage{stmaryrd}
\usepackage{amsfonts}
\usepackage{mathrsfs}
\usepackage{amsmath}
\usepackage{amscd}
\usepackage{graphicx}
\usepackage{booktabs}

\begin{document}
\title{Experimental simulation of anyonic fractional statistics with an NMR quantum information processor}
\author{ Guanru Feng$^{1,2,3}$, Guilu Long$^{1,2}$ and Raymond Laflamme$^{3,4}$}
\address{1 Department of Physics, Tsinghua University, Beijing 100084, China\\
         2 State Key Laboratory of Low-dimensional Quantum Physics, Tsinghua University, Beijing 100084, China\\
         3 Institute for Quantum Computing and Department of Physics, University of Waterloo, Waterloo, Ontario, N2L 3G1, Canada\\
         4 Perimeter Institute for Theoretical Physics, Waterloo, Ontario, N2J 2W9, Canada}
\begin{abstract}
Anyons have exotic statistical properties, fractional statistics, differing from Bosons and Fermions. They can be created as excitations of some Hamiltonian models. Here we present an experimental demonstration of anyonic fractional statistics by simulating a version of the Kitaev spin lattice model proposed by Han et al\cite{PRLDuan} using an NMR quantum information processor. We use a 7-qubit system to prepare a 6-qubit pseudopure state to implement the ground state preparation and realize anyonic manipulations, including creation, braiding and anyon fusion. A $\frac{\pi}{2}\times 2$ phase difference between the states with and without anyon braiding, which is equivalent to two successive particle exchanges, is observed. This is different from the $\pi\times 2$ and $2\pi \times 2$ phases for Fermions and Bosons after two successive particle exchanges, and is consistent with the fractional statistics of anyons.

\end{abstract}
\maketitle

%%%%%%%%%%%%%%%%%%%%%%%%%%%%%%%%%%%%%%%%%%%%%%%%%%%%%%%%%%%%%%%%%%%%%%%%%%%%%%%%%%%%%%%%
\section{INTRODUCTION}

In 3-dimensional space, indistinguishable particles obey Fermi-Dirac statistics (Fermions), or Bose-Einstein statistics (Bosons). For both Fermions and Bosons, upon the exchange of two indistinguishable particles, the system wave function gains a $\pi$ or $2\pi$ phase change.  However, when restricted to 2-dimensional space, particles appear to obey fractional statistics\cite{PhysRevLett.48.1144}. This means that when two indistinguishable particles in 2-dimensional space are exchanged, the system wave function gains a statistical phase change, ranging continuously from $0$ to $2\pi$. Those quasiparticles are defined as anyons.
Anyons can be grouped in abelian and nonabelian anyons. Abelian anyons are particles that realize 1-dimensional representations of braid groups. In nature, abelian anyons are believed to exist and be responsible for the fractional quantum hall effect (FQH)\cite{PhysRevLett.48.1559,PhysRevLett.50.1395,PhysRevLett.53.722}. Nonabelian anyons are particles that behave as multi-dimensional representations of braid groups. They are critical in topological quantum computing, for example, in the Kitaev fault-tolerant quantum computation models \cite{AYu20032,Alexei20062}. Recently, the interest in anyons is enhanced by the developing field of quantum computing because of their potential ability to impliment fault-tolerant quantum computing architecture.

Several theoretical schemes have been proposed to directly observe the fractional statistics associated with the anyon braiding motion\cite{PhysRevLett.94.166802,PhysRevLett.96.016802,PhysRevLett.96.016803,naturep,PhysRevLett.91.090402,nphys287,Zhang20112007,nphys943,PRLDuan,prsa464}. These schemes are mainly grouped into two approaches: the first is proposed to be realized in  FQH systems, and the second makes use of the Kitaev models. In  FQH systems, it is difficult to directly observe anyonic fractional statistics, and to introduce or resolve individual anyons\cite{frometoa} when compared with the schemes using the Kitaev spin lattice models. Experimental demonstrations in photon systems using the Kitaev spin lattice models have been realized\cite{1367-2630-11-8-083010,PhysRevLett.102.030502}. However the anyons are not protected from local noise and there is no explicit particle interpretation of the excitations\cite{nphys943}  because the background Hamiltonian vanishes in such photon systems. In contrast, the background Hamiltonian can be simulated in the nuclear magnetic resonance (NMR) systems.

In our work an NMR quantum information processor is used to demonstrate the anyon braiding scheme proposed by Han et al.\cite{PRLDuan} in the smallest Kitaev system utilizing 6 qubits. The six-body ground state preparation, anyon excitations, and anyonic braiding operations are realized using a 7-qubit molecle in liquid state NMR. By comparing the two final states, of which one is obtained after the anyon creation, braiding, and fusion processes while the other doesn't undergo such processes, the phase difference, which is mapped into a frequency change of NMR spectrum peaks in our experiment, can be observed.

 %%%%%%%%%%%%%%%%%%%%%%%%%%%%%%%%%%%%%%%%%%%%%%%%%%%%%%%%%%%%%%%%%%%%%%%%%%%%%%%%%%%%%%%%%%%%%%%%%%%%%%%
\section{The Kitaev $k\times k$ square lattice model}
\begin{figure}[!ht]
\centering
\includegraphics[width=1.7in,height=1.5in]{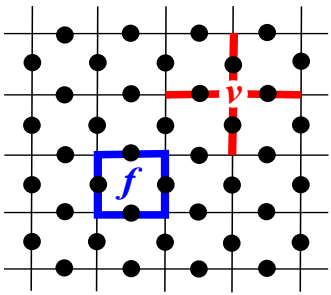}
\caption{Illustration of the first Kitaev model. The first Kitaev spin lattice model is a $k\times k$ square lattice on the torus. There is a qubit on each bond. The operaters $A_{v}$ and $B_{f}$ act on the qubits of the vertex $v$ and face $f$, respectively.}
\label{anyonscheme}
\end{figure}

The first Kitaev spin lattice model\cite{AYu20032} is a $k\times k$ square lattice on the torus, containing qubits on each of the
bonds (Figure \ref{anyonscheme}) (here we define the bonds as the minimal lines forming the lattice). The total number of qubits is $2k^{2}$. The spin lattice contains vertices and faces. A vertex $v$ is the intersection of four bonds. A face $f$ means the surface with boundary  defined by four bonds. We can then define an Hamiltonian as
\begin{align}
H_{K}=-\sum_{v}A_{v}-\sum_{f}B_{f}\label{hamiltonian},
\end{align}
where 
\begin{align}
A_{v}=\Pi_{j\in vertex(v)}\sigma^{x}_{j}, B_{f}=\Pi_{j\in face(f)}\sigma^{z}_{j}
\end{align}
$A_{v}$ ($B_{f}$) represents the 4-body interactions belonging to the qubits which live on the vertex $v$ (or the face $f$). For all the vertices and faces, the ground state $|\Psi _{ground} \rangle$ for the Hamiltonian $H_{K}$ satisfies
\begin{align}
A_{v}|\Psi _{ground} \rangle=|\Psi _{ground} \rangle, B_{f}|\Psi _{ground}\rangle=|\Psi _{ground} \rangle.
\end{align}

The ground states are 4-fold degenerate. They form a protected subspace $G$
\begin{equation}
G=\{|\xi\rangle \in N, A_{v}|\xi\rangle = |\xi\rangle, B_{f}|\xi\rangle = |\xi\rangle, for\ all\ v\ and\ f \}.
\end{equation}
$N$ is the Hilbert space of the $2k^{2}$ qubits. This is the definition of the toric code, which is a special kind of stablizer code\cite{AYu20032}. $A_{v}$ and $B_{f}$ are its stabilizer operators. Because of the periodic boundary conditions, for each qubit $j$, $\sigma^{x}_{j}$ and $\sigma^{z}_{j}$ appear twice in different ${A_{v}}'s$ and ${B_{f}}'s$. It can be easily obtained that
\begin{equation}
\Pi_{v}A_{v}=1, \Pi_{f}B_{f}=1.\label{relationship}
\end{equation}

If a state does not satisfy several (for example, n) of the $A_{v}|\Psi  \rangle=|\Psi \rangle$ and $B_{f}|\Psi\rangle=|\Psi \rangle$ constraints, it's an excited state with n elementary excitations (or quasiparticles). The relationships in Eq. \ref{relationship} implies that the quasiparticles should appear in pairs.

\begin{figure}[!ht]

\includegraphics[width=3in,height=1.6in]{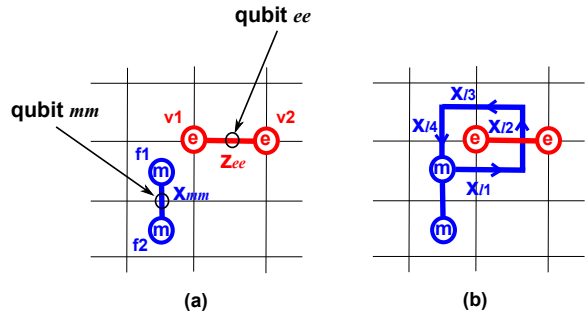}
\caption{Illustration of anyon creation and braiding operations. (a) The creation of two $\it{m}$ ($\it{e}$) anyons by the operation $X_{\it{mm}}$ ($Z_{\it{ee}}$). $X_{\it{mm}}$ means $\sigma_{x}$ operation to qubit $\it{mm}$, and $Z_{\it{ee}}$ means $\sigma_{z}$ operation to qubit $\it{ee}$. The two $\it{m}$ anyons are localized on face $f1$ and face $f2$. The two $\it{e}$ anyons are localized on vertex $v1$ and vertex $v2$. (b) The braiding motion of an $\it{m}$ anyon around an $\it{e}$ anyon by operations $X_{l1}X_{l2}X_{l3}X_{l4}$. Qubits $l1$, $l2$, $l3$, and $l4$ are the qubits along the braiding path.}
\label{anyonbraiding}
\end{figure}

If a $\sigma_{x}$ operation is applied to a qubit, for example, qubit $\it{mm}$, in a ground state, the state wave function is $|\Psi\rangle_{\it{mm}}=X_{\it{mm}}|\Psi _{ground} \rangle$ ($X_{\it{mm}}$ means the $\sigma_{x}$ operation to qubit $\it{mm}$). It satisfies $B_{f1}|\Psi\rangle_{\it{mm}} =-|\Psi\rangle_{\it{mm}}$ and $B_{f2}|\Psi\rangle_{\it{mm}} =-|\Psi\rangle_{\it{mm}}$. $f1$ and $f2$ are the two faces next to qubit $\it{mm}$. That means two quasiparticles have been created at those particular locations. The two quasiparticles can be considered as two ``defects" localized on faces $f1$ and $f2$. They are called $\it{m}$ particles. Instead, if a $\sigma_{z}$ operation is applied to a qubit, for example qubit $\it{ee}$, in a ground state, the state wave function is $|\Psi\rangle_{\it{ee}}=Z_{\it{ee}}|\Psi _{ground} \rangle$ ($Z_{\it{ee}}$ means the $\sigma_{z}$ operation to qubit $\it{ee}$). It satisfies $A_{v1}|\Psi\rangle_{\it{ee}} =-|\Psi\rangle_{\it{ee}}$ and $A_{v2}|\Psi\rangle_{\it{ee}} =-|\Psi\rangle_{\it{ee}}$. This also means two quasiparticles occur. Here, $v1$ and $v2$ are the two neighboring vertices which are connected by the bond qubit $\it{ee}$ lives on. The two quasiparticles can also be considered as ``defects" localized on vertices $v1$ and $v2$. They are called $\it{e}$ particles. The states with quasiparticles (excitations) are excited states. (Figure \ref{anyonbraiding}(a))

Since two $\it{m}$ ($\it{e}$) particles at the same site annihilate, the $\it{m}$ ($\it{e}$) particle can be moved by applying $\sigma_{x}$ ($\sigma_{z}$) operations along the path (Figure \ref{anyonbraiding}(b)). A braiding operation is to move an $\it{m}$ ($\it{e}$) particle around an $\it{e}$ ($\it{m}$) particle along a closed circle path, which is equivalent to two successive particle exchanges\cite{prsa464}. For Fermions and Bosons, states do not change after two successive particle exchanges. For the $\it{m}$ and $\it{e}$ particles, it has been shown that after a braiding operation, the global state gains a $\frac{\pi}{2}\times 2$ phase change\cite{AYu20032}, which is different from Fermions and Bosons. Therefore the $\it{m}$ and $\it{e}$ particles are anyons that obey fractional statistics.

%%%%%%%%%%%%%%%%%%%%%%%%%%%%%%%%%%%%%%%%%%%%%%%%%%%%%%%%%%%%%%%%%%%%%%%%%%%%%%%%%%%%%%%%%%%%%%%%%%%%%%%
\section{The six-qubit Kitaev spin lattice model and the experimental scheme}

\begin{figure}[!ht]
\centering
\includegraphics[width=2in,height=1.5in]{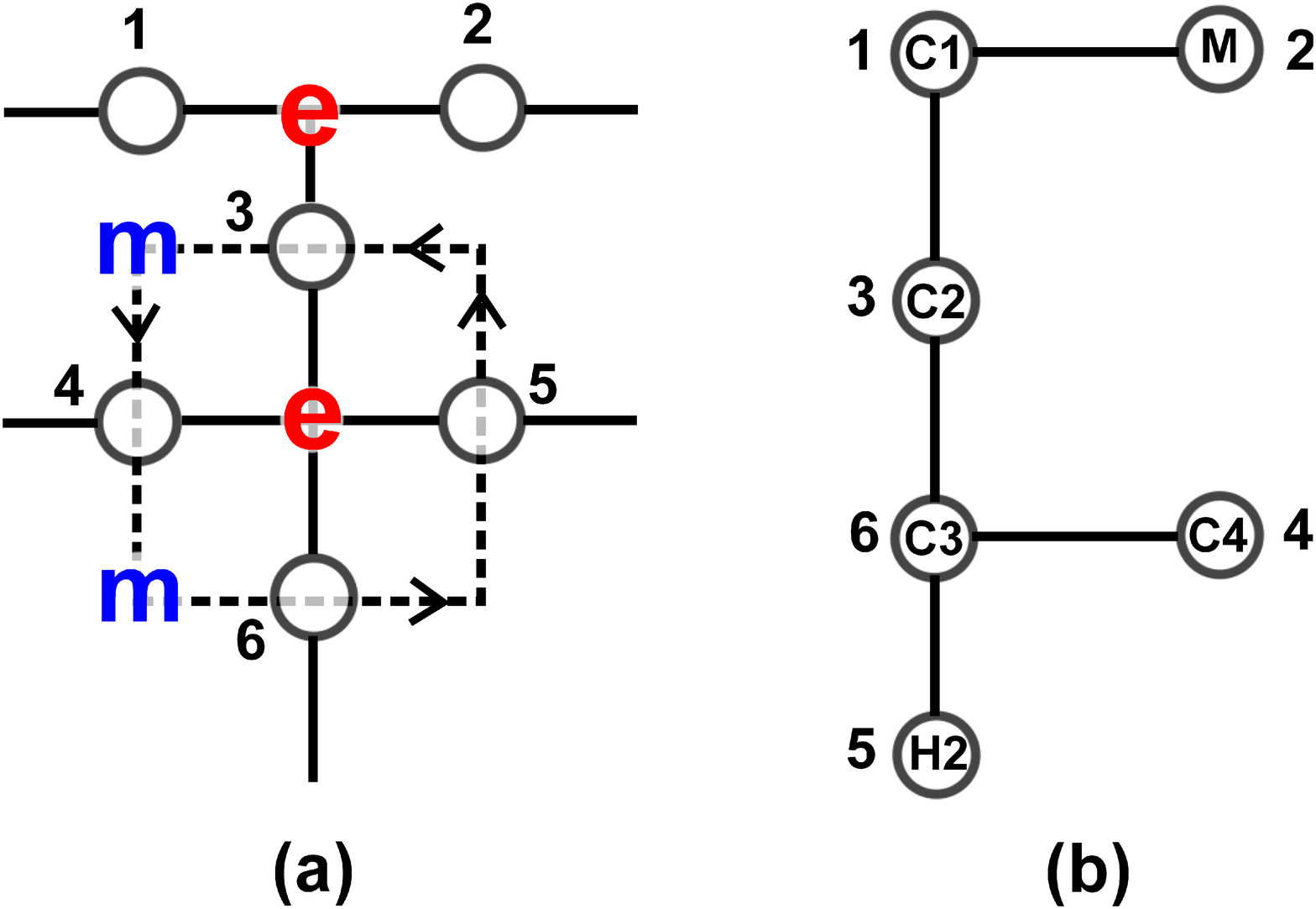}
\caption{(a) The six-qubit Kitaev model and its braiding loop. A pair of $\it{m}$ ($\it{e}$) anyons are created by the operation $X_{4}$ ($Z_{3}$) and the braiding operation is realized by $X_{6}X_{5}X_{3}X_{4}$. (b) The graph state that is equivalent under local unitary operations to the ground state of the Hamiltonian in (a). The corresponding qubits in the NMR spin system are also labeled at each vertex.}
\label{graphicmodel}
\end{figure}

The minimum amount of qubits needed to implement the smallest version of the periodic Kitaev model for anyon braiding operations is eight.
However by abandoning periodic condition, the spin lattice model can be extended from a square lattice to any planar graph\cite{planar}, and an anyonic model can be found with six qubits where we can demonstrate braiding statistics\cite{PRLDuan}. The graphic structure of the six-qubit model is shown in Figure \ref{graphicmodel}(a). The Hamiltonian of the system is
\begin{align}
H_{6}&=-A_{1}-A_2-B_1-B_2-B_3-B_4,
\end{align}
where, 
\begin{align}
A_1&=\sigma^x_1\sigma^x_2\sigma^x_3,\nonumber\\ A_2&=\sigma^x_3\sigma^x_4\sigma^x_5\sigma^x_6,\nonumber\\ B_1&=\sigma^z_1\sigma^z_3\sigma^z_4,\nonumber\\ B_2&=\sigma^z_2\sigma^z_3\sigma^z_5,\nonumber\\
B_3&=\sigma^z_4\sigma^z_6,\nonumber\\
B_4&=\sigma^z_5\sigma^z_6.\nonumber
\end{align} 

The ground state of the six-qubit Kitaev spin lattice is \begin{align}|\Psi _{ground}\rangle=\frac{1}{2} (|000000\rangle + |111000\rangle + |110111\rangle + |001111\rangle).\end{align} Because the boundary conditions have changed, the ground state is not degenerate anymore. The ground state can be created from a six-qubit graph state shown in Figure \ref{graphicmodel}(b)\cite{PRLDuan}. A graph state is a type of multi-qubit state represented by a graph with the vertex set $V$ and the edge set $E$, and is defined as \begin{align}|G\rangle=\Pi_{(i,j)\in E}U_{i,j}|+\rangle^{\bigotimes V},\end{align} where the operater $U_{i,j}$ is the controlled-$\sigma_z$ operation between the qubits $i$ and $j$, and $|+\rangle=\frac{1}{\sqrt{2}}(|0\rangle +|1\rangle)$. The six-qubit graph state corresponding to the graph in Figure \ref{graphicmodel}(b) is \begin{align}|G_{6}\rangle=U_{1,2}U_{1,3}U_{3,6}U_{4,6}U_{5,6}|++++++\rangle.\end{align}
The ground state of the six-qubit Kitaev spin lattice is \begin{align}|\Psi _{ground}\rangle=O|G_{6}\rangle,\end{align} where $O=IHHHHI$, $I$ is the identity operater, and $H$ is the Hadamard operater. The $|\Psi _{ground}\rangle$ can be prepared by first preparing a six-qubit graph state $|G_{6}\rangle$, then implementing an $O$ operation. This gives a method to prepare the ground state.

For the ground state of the six-qubit system in Figure \ref{graphicmodel}(a), if a $\sigma_x$ operation is applied to qubit 4, a pair of $\it{m}$ particles are created on its two neighboring faces, and if a $\sigma_z$ is applied to qubit 3, a pair of $\it{e}$ particles are created on its two neighboring vertices. By applying successive $\sigma_x$ operations on qubits 6, 5, 3, and 4, one $\it{m}$ particle moves in a loop around one $\it{e}$ particle. After such a braiding operation, the global wave function will obtain a phase factor of  -1. By observing this phase change, one can verify the fractional statistics of anyons.

Han et al.\cite{PRLDuan} give the basic circuit for ground state preparation, anyon creation, anyon braiding, and anyon fusion. It should be noted that if one does the braiding operation to a state with a pair of $\it{e}$ particles and a pair of $\it{m}$ particles (Figure \ref{anyonbraiding}(b)), the phase change after braiding is a global phase, which cannot be observed in experiments directly. In the scheme proposed by Han et al.\cite{PRLDuan}, the anyon creation step is realize by $\sigma_x$ and $\sqrt{\sigma_z}=e^{i\frac{\pi}{4}}e^{-i\frac{\pi}{4}\sigma_{z}}$ instead of $\sigma_x$ and $\sigma_z$. The $\sigma_x$ operation creates a pair of $\it{m}$ particles. The $\sqrt{\sigma_z}$ operation creates a superposition between the states with and without a pair of $\it{e}$ particles. Therefore after the anyon creation step, the state of the system is \begin{align}|\Psi\rangle=\frac{1}{\sqrt{2}}(|\psi_{1}\rangle+|\psi_{2}\rangle),\end{align} where $|\psi_{1}\rangle$ is a state with a pair of $\it{m}$ particles only, $|\psi_{2}\rangle$ is a state with a pair of $\it{m}$ particles together with a pair of $\it{e}$ particles. $|\psi_{1}\rangle$ will not change after a braiding procedure, because no $\it{e}$ particles exist. $|\psi_{2}\rangle$ will obtain a phase factor of -1 after a braiding procedure. Therefore the total wave fuction becomes \begin{align}|\Psi^{'}\rangle=\frac{1}{\sqrt{2}}(|\psi_{1}\rangle-|\psi_{2}\rangle).\end{align} In this way, the phase change caused by braiding operation becomes a local phase factor in front of $|\psi _{2} \rangle$ and is observable in experiments.

The fusion operation is realized by applying $\sqrt{\sigma_z}$ and $\sigma_x$. In such case, if indeed there is a statistical phase change $\frac{\pi}{2}\times 2$ acquired after braiding, the state after fusion is $|\Psi_{ground}\rangle$. This means the $\it{e}$ particle pair and the $\it{m}$ particle pair are both fused. 

In our experiment scheme, $\sqrt{\sigma_z}^{-1}$ and $\sigma_x$ operations are performed as the fusion step (Figure \ref{curcuits}). With the statistical phase change $\frac{\pi}{2}\times 2$ introduced by a braiding operation (two successive exchanges between anyons), the state after the fusion step should be $|\Psi_{excited}\rangle=\sigma_z |\Psi_{ground}\rangle$. Without the statistical phase change $\frac{\pi}{2}\times 2$, the state after the fusion step should be $|\Psi_{ground}\rangle$. Therefore, by observing the difference between the state after the fusion step and the ground state, we can demonstrate the fractional statistics of anyons.

%%%%%%%%%%%%%%%%%%%%%%%%%%%%%%%%%%%%%%%%%%%%%%%%%%%%%%%%%%%%%%%%%%%%%%%%%%%%%%%%%%%%%%%%%%%%%%%%%%%%%%%
\section{Experimental implementation}

\begin{figure}[!ht]
\centering
\includegraphics[width=3in,height=2.5in]{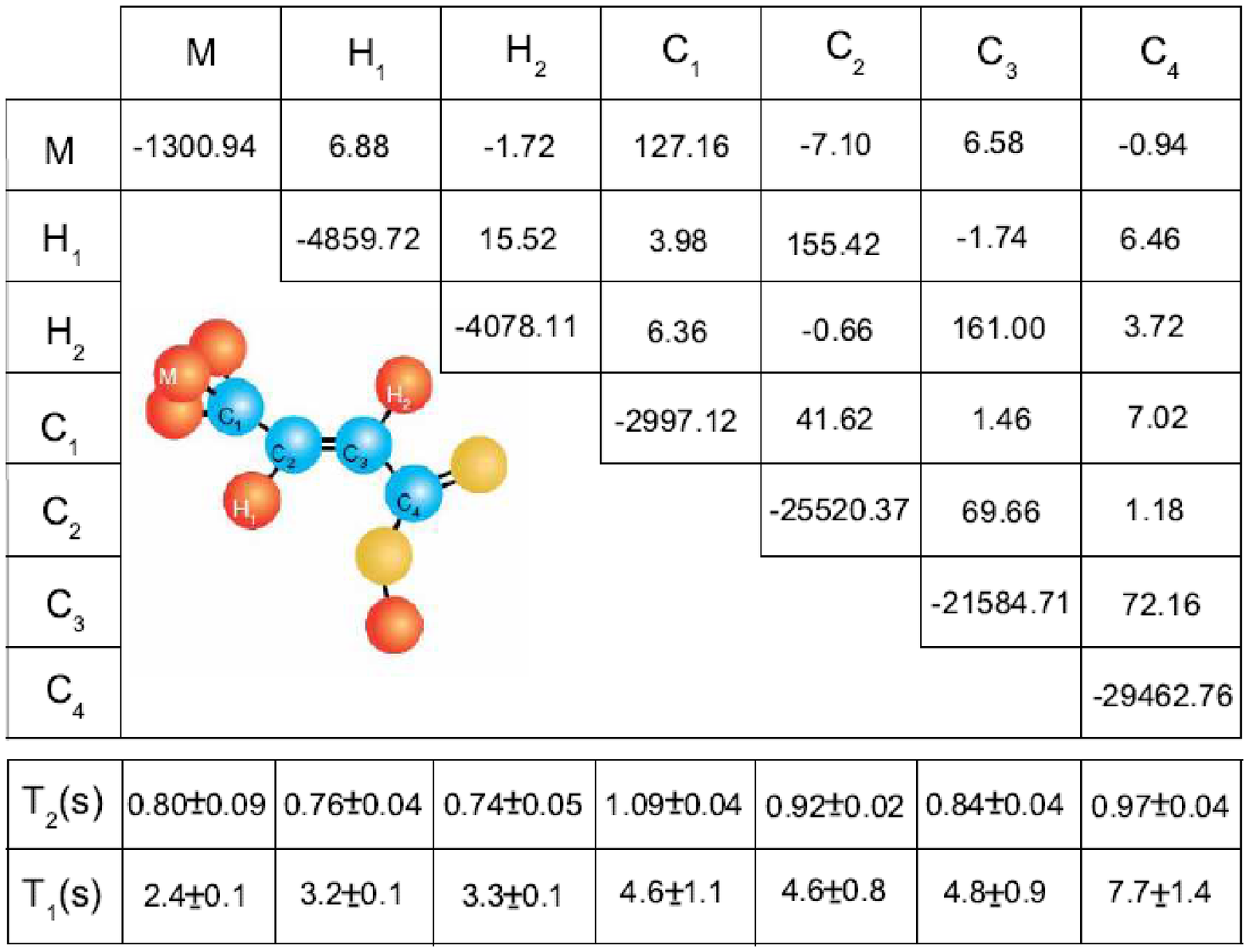}
\caption{Characteristics of the molecule of trans-crotonic acid\cite{nature,molecule}. The chemical shifts (diagonal elements) and J-coupling constants (off-diagonal elements) are given in $Hz$. The spin-lattice and spin-spin relaxation times T1 and T2 are listed at the bottom. The chemical shifts are given with respect to reference frequencies of 700.13 $MHz$ (hydrogens) and
176.05 $MHz$ (carbons). The three hydrogen nuclei in the methyl form a spin-$3/2$ group M. After using a gradient-based subspace selection, this group acts in its spin-$1/2$ subspace\cite{nature}. Therefore, M can be used as a single qubit. We thus have 7 qubits, including the two protons and four carbons.}
\label{moleculestructure}
\end{figure}

In the experiment, $^{13}C$-labeled trans-crotonic acid dissolved in d6 acetone was used. The system contains 7 qubits. The molecule and its parameters are described in Figure \ref{moleculestructure}. 

To implement the anyon braiding, we first prepared the molecule in the labeled pseudopure state with a deviation matrix of the form  
$\rho_i=Z_{H1}{\bf{0}}_{C1}{\bf{0}}_{M}{\bf{0}}_{C2}{\bf{0}}_{C4}{\bf{0}}_{H2}{\bf{0}}_{C3}$ where
$\bf{0}=|0\rangle\langle 0|$ and $Z$ is the Pauli matrix $\sigma_z$.
 We chose H1 as the label qubit, and C1, M, C2, C4, H2, C3 as qubits 1, 2, 3, 4, 5 and 6, to match to Figure \ref{graphicmodel}(b), using the neighboring couplings for shortening  the gate operations in implementation.

\begin{figure}[!ht]
\centering
\includegraphics[width=3.5in,height=1.6in]{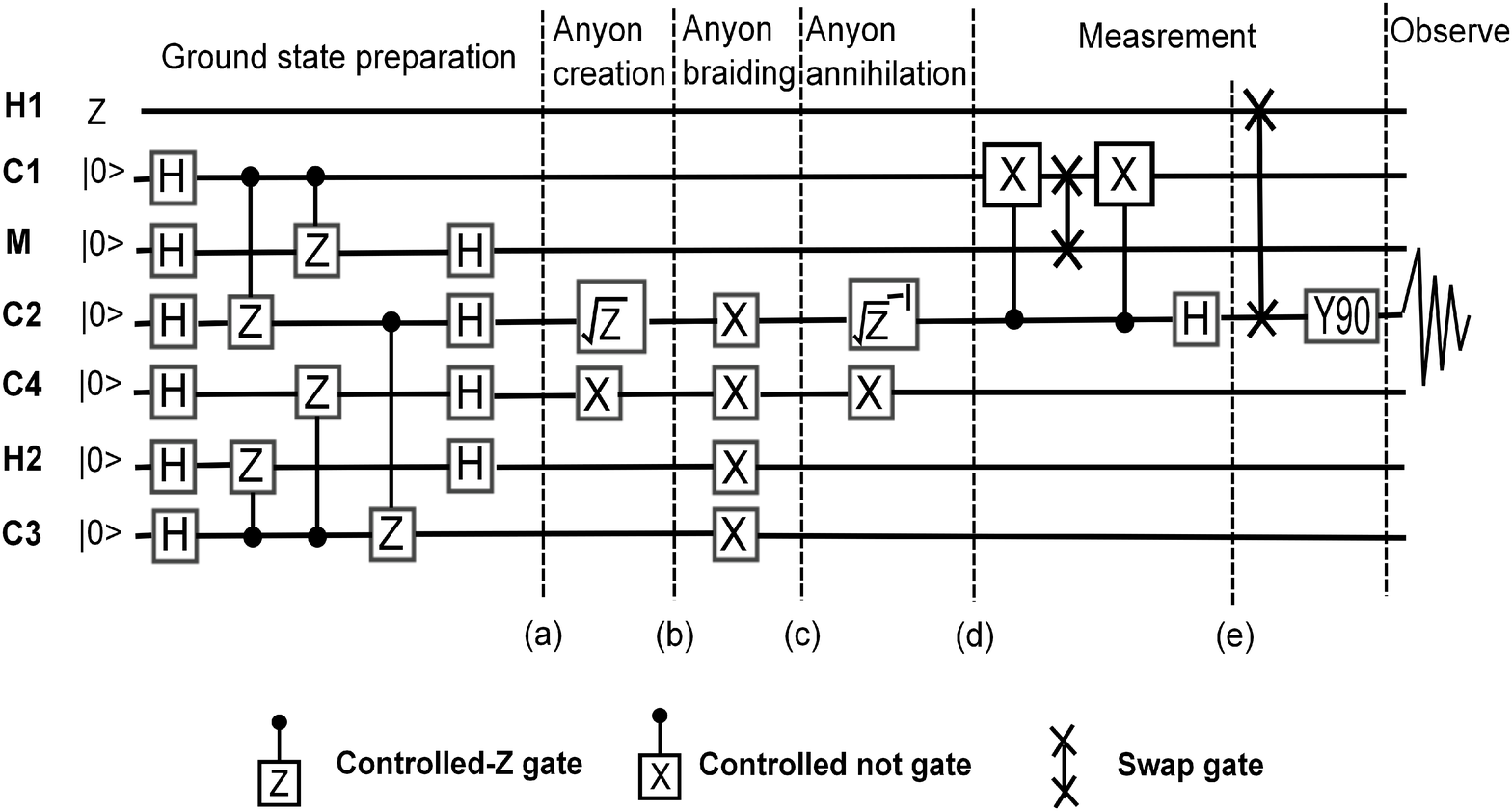}
\caption{The quantum network for the experiment with anyonic manipulation. $H$ represents the Hadamard operation. $Z=\sigma_z$, $X=\sigma_x$. $Y90$ is the read pulse, $Y90=e^{-i\frac{\pi}{4}\sigma_{y}}$.}
\label{curcuits}
\end{figure}

\begin{figure}[!ht]
\centering
\includegraphics[width=2.8in,height=1.6in]{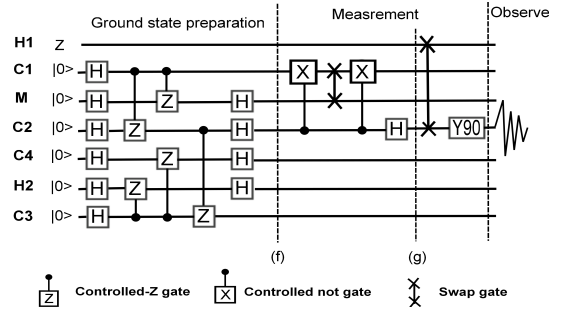}
\caption{The quantum network for the experiment without anyonic manipulation. It serves as a comparison with the experiment in Figure \ref{curcuits}.}
\label{curcuiti}
\end{figure}

Figure \ref{curcuits} and Figure \ref{curcuiti} show the circuits for the experiments with and without anyonic manipulation, respectively. There is a part indicated as ``Measurement"  from which one can observe the phase change from the spectrum of the label qubit. These two experiments illustrated in Figure \ref{curcuits} and Figure \ref{curcuiti} were carried out for comparison.

The wave fuctions of the labeled states (a), (b), (c), (d), (e) in Figure \ref{curcuits}, and (f), (g) in Figure \ref{curcuiti} are as follows:
\begin{align}
|\Psi _a \rangle&=\frac{1}{2} (|0_{C1}0_{M}0_{C2}0_{C4}0_{H2}0_{C3}\rangle + |1_{C1}1_{M}1_{C2}0_{C4}0_{H2}0_{C3}\rangle \nonumber\\&+ |1_{C1}1_{M}0_{C2}1_{C4}1_{H2}1_{C3}\rangle + |0_{C1}0_{M}1_{C2}1_{C4}1_{H2}1_{C3}\rangle)\nonumber\\&=|\Psi _{ground} \rangle\nonumber
\end{align}
\begin{align}
|\Psi _b \rangle&=\frac{1}{2} (|0_{C1}0_{M}0_{C2}1_{C4}0_{H2}0_{C3}\rangle + i|1_{C1}1_{M}1_{C2}1_{C4}0_{H2}0_{C3}\rangle \nonumber\\&+ |1_{C1}1_{M}0_{C2}0_{C4}1_{H2}1_{C3}\rangle + i|0_{C1}0_{M}1_{C2}0_{C4}1_{H2}1_{C3}\rangle)\nonumber\\&=X_{C4}(\frac{1}{\sqrt{2}}(e^{i\frac{\pi}{4}}|\Psi _{ground} \rangle+e^{-i\frac{\pi}{4}}|\Psi _{excited} \rangle))\nonumber\\&=\frac{1}{\sqrt{2}}(|\psi_{1}\rangle+|\psi_{2}\rangle)\nonumber
\end{align}

\begin{align}
|\Psi _c \rangle&=\frac{1}{2} (|0_{C1}0_{M}1_{C2}0_{C4}1_{H2}1_{C3}\rangle + i|1_{C1}1_{M}0_{C2}0_{C4}1_{H2}1_{C3}\rangle \nonumber\\&+ |1_{C1}1_{M}1_{C2}1_{C4}0_{H2}0_{C3}\rangle + i|0_{C1}0_{M}0_{C2}1_{C4}0_{H2}0_{C3}\rangle)\nonumber\\&=X_{C4}(\frac{1}{\sqrt{2}}(e^{i\frac{\pi}{4}}|\Psi _{ground} \rangle-e^{-i\frac{\pi}{4}}|\Psi _{excited} \rangle))\nonumber\\&=\frac{1}{\sqrt{2}}(|\psi_{1}\rangle-|\psi_{2}\rangle)\nonumber
\end{align}
\begin{align}
|\Psi _d \rangle&=\frac{i}{2} (|0_{C1}0_{M}0_{C2}0_{C4}0_{H2}0_{C3}\rangle - |1_{C1}1_{M}1_{C2}0_{C4}0_{H2}0_{C3}\rangle \nonumber\\&+ |1_{C1}1_{M}0_{C2}1_{C4}1_{H2}1_{C3}\rangle - |0_{C1}0_{M}1_{C2}1_{C4}1_{H2}1_{C3}\rangle)\nonumber\\&=iZ_{C2}|\Psi _{ground}\rangle=i|\Psi _{excited} \rangle \nonumber
\end{align}
\begin{align}
|\Psi _e \rangle&=\nonumber\\&\frac{i\sqrt{2}}{2} (|0_{C1}0_{M}1_{C2}0_{C4}0_{H2}0_{C3}\rangle + |1_{C1}1_{M}1_{C2}1_{C4}1_{H2}1_{C3}\rangle)\nonumber
\end{align}
\begin{align}
|\Psi _f \rangle&=\frac{1}{2} (|0_{C1}0_{M}0_{C2}0_{C4}0_{H2}0_{C3}\rangle + |1_{C1}1_{M}1_{C2}0_{C4}0_{H2}0_{C3}\rangle \nonumber\\&+ |1_{C1}1_{M}0_{C2}1_{C4}1_{H2}1_{C3}\rangle + |0_{C1}0_{M}1_{C2}1_{C4}1_{H2}1_{C3}\rangle)\nonumber\\&=|\Psi _{ground} \rangle\nonumber
\end{align}
\begin{align}
 |\Psi _g \rangle&=\nonumber\\&\frac{\sqrt{2}}{2} (|0_{C1}0_{M}0_{C2}0_{C4}0_{H2}0_{C3}\rangle + |1_{C1}1_{M}0_{C2}1_{C4}1_{H2}1_{C3}\rangle)
\end{align}

In the experiment with anyonic manipulation, $|\Psi _a \rangle$ is obtained after the ground state preparation. It is the ground state of the six-qubit Kitaev model. After anyon creation, the wave function of the state is $|\Psi _b \rangle$. After anyon braiding, $|\Psi _b \rangle$ is transformed to $|\Psi _c \rangle$. $|\Psi _b \rangle$ and $|\Psi _c \rangle$ are the superpositions of $|\psi_{1}\rangle$ and $|\psi_{2}\rangle$. $|\psi_{1}\rangle$ is a state with a pair of $\it{m}$ particles. $|\psi_{2}\rangle$ is a state with a pair of $\it{e}$ particles and a pair of $\it{m}$ particles. Due to the $\it{e}$ anyons present in $|\psi_{2}\rangle$, it has a phase change acquired when the $\it{m}$ particle braiding around the $\it{e}$ particle. This causes the difference between $|\Psi _b \rangle$ and $|\Psi _c \rangle$. After anyon annihilation, the state goes to the excited state $|\Psi _d \rangle$. In order to observe the phase change in relatively simple spectra directly, the state $|\Psi _d \rangle$ is transformed to $|\Psi _e \rangle$. For the state $|\Psi _e \rangle$, by observing the label qubit, a 2-peak spectrum can be obtained. The left peak in the spectrum corresponds to $|111111\rangle$ while the right peak corresponds to $|001000\rangle$. 

In the experiment without anyonic manipulation, the ground state preparation  and the measurement procedures are the same as the experiment with anyonic manipulation. The state $|\Psi _f \rangle$ equals $|\Psi _a \rangle$. The state $|\Psi _g \rangle$ is almost the same as $|\Psi _e \rangle$, except that the qubit C2 is $|0\rangle$ in $|\Psi _g \rangle$ (the global phase is ignored). For the state $|\Psi _g \rangle$, by observing the label qubit, a 2-peak spectrum can be obtained. The left peak in the spectrum corresponds to $|110111\rangle$ while the right peak corresponds to $|000000\rangle$. It is straightforward to conclude that the difference is caused by anyon braiding, illustrated in Figure \ref{curcuits}. Because $\it{m}$ and $\it{e}$ anyons do not obey integral statistics, after two successive exchangings they obtain a phase factor, which is mapped into a frequency change of peaks in our experiment.

At the end of the measurement part, the states of qubits H1 and C2 are exchanged via a swap gate. Therefore, C2 becomes the label qubit. The final results are obtained by implementing a $\pi/2$ read pulse to C2. It should be mentioned that the J-coupling constant between C2 and H2 $|J_{C2,H2}|=0.66Hz$ is the smallest of the couplings between C2 and the other nuclei. $|J_{C2,H2}|$ can be resolved in our experiments, which means all the 64 peaks of C2 spectrum are resolved (Figure \ref{fitting}). $|J_{C2,H1}|=155.42Hz$ is the largest J-coupling constant of C2. The braiding induced changes in frequencies of peaks should equal $|J_{C2,H1}|$, which can be easily and explicitly observed in the experiments.

The gates used in the ground state preparation were realized by combining single-qubit rotations and evolutions of the J-coupling constants between the neighboring qubits, while all the anyonic manipulations were realized by single qubit rotations\cite{PhysRevA.52.3457,apl76}. The single-qubit rotation pulses were generated using the GRAPE algorithm\cite{GRAPE} for H1 and H2, and were standard Isech-shaped r.f. pulses for M and C1-C4. The J-coupling evolutions were realized by implementing refocusing pulses. We combined all the pulses using a custom-built software compiler, which numerically optimizes refocusing pulses and  minimizes the errors due to undesired J-coupling evolutions\cite{natcomm,PhysRevA.78.012328}. The duration of the pulse sequences shown in Figure \ref{curcuiti} and Figure \ref{curcuits} was 195.1ms and 250.2ms, respectively.

\begin{widetext}
\begin{center}

\begin{figure}[!ht]
\includegraphics[width=5in,height=2.5in]{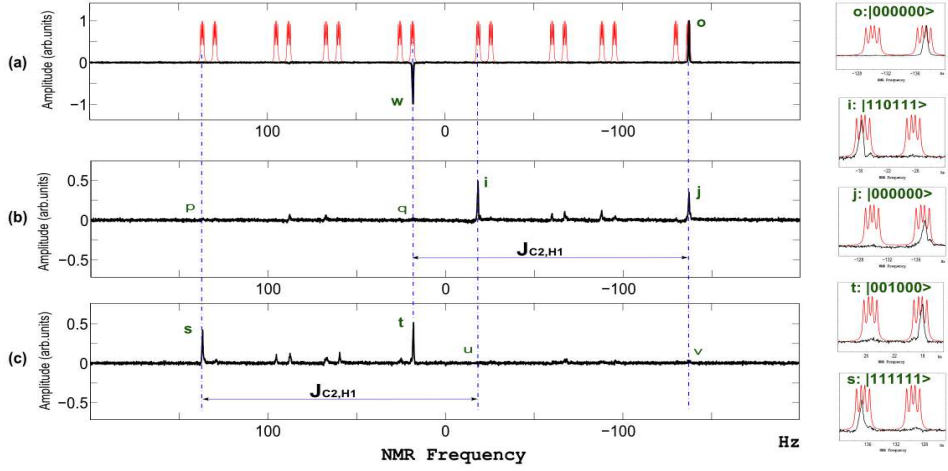}
\caption{(a) The superposed spectra of the theoretical C2 thermal state spectrum (red line) and the experimental pseudopure state spectrum (black line). There are 64 peaks in the thermal state spectrum, each corresponding to a computational basis state. (b) The experimental spectrum corresponding to the experiment in Figure \ref{curcuiti}. It has two dominant peaks, i and j. (c) The experimental spectrum corresponding to the experiment in Figure \ref{curcuits}. It has two dominant peaks, s and t. The amplitude of peak o in the experimental pseudopure state spectrum in (a) is taken as reference to normalize the experimental signals shown in (b) and (c). On the right are the zoomed-in spectra for the peaks o, i, j, t, and s. The states which the experimental peaks correspond to are labeled on top of each zoomed-in spectrum. There is a $J_{C2,H1}$ distance between peaks i and s, j and t. p, q, u, and v are the small peaks that have the same frequencies as those of peaks s, t, i, and j, respectively.}
\label{fitting}
\end{figure}
\end{center}
\end{widetext}

%%%%%%%%%%%%%%%%%%%%%%%%%%%%%%%%%%%%%%%%%%%%%%%%%%%%%%%%%%%%%%%%%%%%%%%%%%%%%%%%%%%%%%%%%%%%%%%%%%%%%%%
\section{Experimental results}

The final results for the experiments are shown in Figure \ref{fitting}. Figure \ref{fitting}(a) shows the superposed spectra between the simulated thermal state spectrum of C2 and the experimental pseudopure state spectrum. It shows the experimentally realized $|000000\rangle$ peak (peak o). It should be mentioned that there is an antiphase peak (peak w) in Figure \ref{fitting}(a). This antiphase peak was caused by the label qubit H1, which was in the $Z$ state. Figure \ref{fitting}(b) displays the spectrum of the state $\frac{i\sqrt{2}}{2} (|0_{C1}0_{M}0_{H1}0_{C4}0_{H2}0_{C3}\rangle + |1_{C1}1_{M}0_{H1}1_{C4}1_{H2}1_{C3}\rangle)$, and Figure \ref{fitting}(c) displays the spectrum of the state \\$\frac{i\sqrt{2}}{2} (|0_{C1}0_{M}1_{H1}0_{C4}0_{H2}0_{C3}\rangle + |1_{C1}1_{M}1_{H1}1_{C4}1_{H2}1_{C3}\rangle)$, both generated by observing C2, whose deviation density matrix was $X=\sigma_{x}$. Peaks i (-18.3$Hz$), j (-137.2$Hz$), s (137.1$Hz$) and t (18.2$Hz$) correspond to states  $|110111\rangle$, $|000000\rangle$, $|111111\rangle$ and $|001000\rangle$, respectively. The sum intensity of the two dominant peaks, both in Figure \ref{fitting}(b) and Figure \ref{fitting}(c), is about 0.7, normalized using the intensity of peak o.

There is a $155.4Hz=|J_{C2,H1}|$ shift between peaks i and s, j and t (Figure \ref{fitting}). This frequency difference between the peaks in the two spectra was caused by the process of anyonic manipulation, demonstrating that after the braiding operation, the state with $\it{e}$ and $\it{m}$ anyons acquired a phase change $\delta=(\frac{\pi}{2} + \eta)*2$. Here $2\eta$ is the deviation of the experimental phase change from $\frac{\pi}{2}\times 2$.

We suppose the experimentally realized labeled state (a) in Figure \ref{curcuits} and labeled state (f) in Figure \ref{curcuiti} were
\begin{align}
|{\Psi _f}' \rangle=|{\Psi _a}' \rangle=\alpha |\Psi _{ground} \rangle +\beta |\Psi _{excited} \rangle + \gamma |\Psi _{error} \rangle.
\end{align} 
The above expression implies the ground state preparation was not perfect. $\beta |\Psi _{excited} \rangle$ was responsible for peaks p and q. $\gamma |\Psi _{error} \rangle$ was responsible for the peaks other than peaks i, j, p and q. (Figure \ref{fitting}(b)) The experimentally realized labeled state (g) in Figure \ref{curcuiti} was 
\begin{align}
|{\Psi _g}' \rangle=\frac{\sqrt{2}}{2} [\alpha(&|0_{C1}0_{M}0_{C2}0_{C4}0_{H2}0_{C3}\rangle \nonumber\\  + &|1_{C1}1_{M}0_{C2}1_{C4}1_{H2}1_{C3}\rangle)\nonumber\\ +\beta(&|0_{C1}0_{M}1_{C2}0_{C4}0_{H2}0_{C3}\rangle \nonumber\\ + &|1_{C1}1_{M}1_{C2}1_{C4}1_{H2}1_{C3}\rangle)]+\gamma |{\Psi _{error}}' \rangle.
\end{align}
$|\frac{\beta}{\alpha}|$ can be determined from the peak intensities (denoted as $\Gamma$) of the experimental spectrum shown in Figure \ref{fitting}(b). 
\begin{align}|\frac{\beta}{\alpha}|=\sqrt{\frac{\Gamma _{p}+\Gamma _{q}}{\Gamma _{i}+\Gamma _{j}}}=0.18\pm 0.09. \label{grounddata}
\end{align} 

The experimentally realized labeled states (d) and (e) in Figure \ref{curcuits} were
\begin{align}
|{\Psi _d}' \rangle&=ie^{i\eta}[(-\alpha \sin{\eta} -\beta \cos{\eta}) |\Psi _{ground} \rangle\nonumber\\ &+(\alpha \cos{\eta}-\beta \sin{\eta})|\Psi _{excited} \rangle ]+ \gamma |{\phi _{error}} \rangle\nonumber\\ &=ie^{i\eta}[\alpha '|\Psi _{ground} \rangle+\beta '|\Psi _{excited} \rangle ]+ \gamma |{\phi _{error}}\rangle.\label{stated}
\end{align}
 
\begin{align}
|{\Psi _e}' \rangle=\frac{\sqrt{2}ie^{i\eta}}{2} &[\alpha ' (|0_{C1}0_{M}0_{C2}0_{C4}0_{H2}0_{C3}\rangle \nonumber\\  &+ |1_{C1}1_{M}0_{C2}1_{C4}1_{H2}1_{C3}\rangle)\nonumber\\ &+\beta ' (|0_{C1}0_{M}1_{C2}0_{C4}0_{H2}0_{C3}\rangle \nonumber\\ &+ |1_{C1}1_{M}1_{C2}1_{C4}1_{H2}1_{C3}\rangle)]+\gamma |{\phi _{error}}' \rangle.
\end{align}
Here $\eta=\frac{\delta}{2}-\frac{\pi}{2}$, $\alpha '=-\alpha \sin{\eta} -\beta \cos{\eta}$, $\beta '=\alpha \cos{\eta}-\beta \sin{\eta}$. $\alpha '|\Psi _{ground} \rangle$ was responsible for peaks u and v. $\gamma |{\phi _{error}} \rangle$ was transformed from $\gamma |\Psi _{error} \rangle$ via anyonic manipulation. $\gamma |{\phi _{error}} \rangle$ was responsible for the peaks other than s, t, u and v. (Figure \ref{fitting}(c)) 

$|\frac{\beta '}{\alpha '}|$ can be determined from the experimental spectrum shown in Figure \ref{fitting}(c). 
\begin{align}
|\frac{\alpha '}{\beta '}|=\sqrt{\frac{\Gamma _{u}+\Gamma _{v}}{\Gamma _{s}+\Gamma _{t}}}=0.24\pm 0.06. \label{exciteddata}
\end{align}
Combining Equation \ref{grounddata} and Equation \ref{exciteddata}, we can obtain 
\begin{align}
\tan{\eta}=\frac{|\frac{\alpha '}{\beta '}|-|\frac{\beta}{\alpha}|}{1+|\frac{\beta}{\alpha}|*|\frac{\alpha '}{\beta '}|}=0.06 \pm 0.03.
\end{align}

$\eta =0.06\pm 0.03=(0.02\pm 0.01)\pi$ and the phase change $\delta=(\frac{\pi}{2} + \eta)*2=(0.52\pm 0.01)\pi\times 2$. This agrees with the prediction of the fractional statistics.\\

\section{Discussion}

The signal loss mainly came from the spin-spin relaxation and pulse imperfection. Through comparing the signal intensities of the simulation with and without the T2 effects, we estimate that T2 effects contributed to about 15\% of the loss of signal. Comparing the experimental results (peak intensity $\sim$0.7) with the results of the simulation with T2 effects (peak intensity $\sim$1), we estimate that imperfections of the implementation of rf pulses caused an additional approximate 15\% signal loss.

The values of $\eta$ in the simulation with and without T2 effects are $0.07$ and $0.06$, respectively, which match well with the experimental data ($\eta =0.06\pm 0.03$). The difference between the two $\eta$ values in the simulation is small, which means decoherence did not contribute much to the deviation of $\delta$ from $\frac{\pi}{2}\times 2$. The deviation was mainly caused by imperfections of the refocusing protocols and the implementation of rf pulses.

%%%%%%%%%%%%%%%%%%%%%%%%%%%%%%%%%%%%%%%%%%%%%%%%%%%%%%%%%%%%%%%%%%%%%%%%%%%%%%%%%%%%%%%%%%%%%%%%%%%%%%%
\section{Conclusion}
In summary, we demonstrated the anyonic fractional statistics, using a 7-qubit NMR system. This is the first demonstration on topological quantum computing using nuclear spins. One advantage of our experimental scheme is that we can use the same technique to simulate a 9-qubit Kitaev spin model mentioned by Han et al.\cite{PRLDuan} using an NMR system with more spins, so that we can do a demonstration that anyonic operation is robust to different braiding
paths and thus taking an important step  towards showing  the fault tolerance  properties of the Kitaev model.\\

\section*{ACKNOWLEDGMENTS}
We thank J.-F. Zhang for helpful discussions, and Industry Canada for support at the Institute for Quantum Computing. R.L. acknowledges support from CIFAR and NSERC. G.L. acknowledges support from the National Natural Science Foundation of China (Grant No. 10874098), and the National Basic Research Program of China
(2009CB929402).

%%%%%%%%%%%%%%%%%%%%%%%%%%%%%%%%%%%%%%%%%%%%%%%%%%%%%%%%%%%%%%%%%%%%%%%%%%%%%%%%%%%%%%%%%%%%%%%%%%%%%%%


\begin{thebibliography}\item
\bibitem{PRLDuan}Y.-J. Han, R. Raussendorf, and L.-M. Duan, Phys. Rev. Lett. 98, 150404 (2007).

\bibitem{PhysRevLett.48.1144}F. Wilczek, Phys. Rev. Lett. 48, 1144 (1982).
\bibitem{PhysRevLett.48.1559}D. C. Tsui, H. L. Stormer, and A. C. Gossard, Phys. Rev. Lett. 48, 1559 (1982).
\bibitem{PhysRevLett.50.1395}R. B. Laughlin, Phys. Rev. Lett. 50, 1395 (1983).
\bibitem{PhysRevLett.53.722}D. Arovas, J. R. Schrieffer, and F. Wilczek, Phys. Rev. Lett. 53, 722 (1984).
\bibitem{AYu20032}A. Yu. Kitaev, Annals of Physics 303, 2 (2003).
\bibitem{Alexei20062}A. Yu. Kitaev, Annals of Physics 321, 2 (2006).
\bibitem{PhysRevLett.94.166802}S. Das Sarma, M. Freedman, and C. Nayak, Phys. Rev. Lett. 94, 166802 (2005).
\bibitem{PhysRevLett.96.016802}A. Stern, and B. I. Halperin, Phys. Rev. Lett. 96, 016802(2006).
\bibitem{PhysRevLett.96.016803}P. Bonderson, A. Kitaev, and K. Shtengel, Phys. Rev. Lett. 96, 016803 (2006).
\bibitem{naturep}C. Weeks, G. Rosenberg, B. Seradjeh, and M. Franz, Nat. Phys. 3, 796 (2007).
\bibitem{PhysRevLett.91.090402}L.-M. Duan, E. Demler, and M. D. Lukin, Phys. Rev. Lett. 91 090402 (2003).
\bibitem{nphys287}A. Micheli, G. K. Brennen, and P. Zoller, Nat. Phys. 2, 341 (2006).
\bibitem{Zhang20112007}C. Zhang, V. W. Scarola, S. Tewari, and S. Das Sarma, Proceedings of the National Academy of Sciences 104, 18415 (2007).
\bibitem{nphys943}L. Jiang, G. K. Brennen, A. V. Gorshkov, K. Hammerer, M. Hafezi, E. Demler, M. D. Lukin, and P. Zoller, Nat. Phys. 4, 482 (2008).
\bibitem{frometoa}F. Wilczek, Phys. World 19, 22 (2006).
\bibitem{1367-2630-11-8-083010}J. K. Pachos, W. Wieczorek, C. Schmid, N. Kiesel, R. Pohlner, and H. Weinfurter, New Journal of Physics 11, 083010 (2009).
\bibitem{PhysRevLett.102.030502}C.-Y. Lu, W.-B. Gao, O. G\"uhne, X.-Q. Zhou, Z.-B. Chen, and J.-W. Pan, Phys. Rev. Lett. 102, 030502(2009).

\bibitem{prsa464}G. K. Brennen, and J. K. Pachos, Proceedings of the Royal Society A: Mathematical, Physical and Engineering Sciences 464, 1 (2008).

\bibitem{planar}E. Dennis, A. Kitaev, A. Landahl, and J. Preskill, J. Math. Phys. 43, 4452 (2002).

\bibitem{nature}E. Knill, R. Laflamme, R. Martinez, and C.-H. Tseng, Nature 404, 368 (2000).
\bibitem{molecule}J.-F. Zhang, M. Grassl, B. Zeng, and R. Laflamme, Phys. Rev. A 85, 062312 (2012).


\bibitem{Chuang}I. L. Chuang, N. Gershenfeld, M. G. Kubinec, and D. W. Leung, Proceedings of the Royal Society of London, Series A: Mathematical, Physical and Engineering Sciences 454, 447 (1998).
\bibitem{PhysRevA.52.3457}A. Barenco, C. H. Bennett, R. Cleve, D. P. DiVincenzo, N. Margolus, P. Shor, T. Sleator, J. A. Smolin, and H. Weinfurter, Phys. Rev. A 52, 3457 (1995).
\bibitem{apl76}L. M. K. Vandersypen, M. Steffen, M. H. Sherwood, C. S. Yannoni, G. Breyta, and I. L. Chuang, Applied Physics Letters 76, 646 (2000).
\bibitem{GRAPE}N. Khaneja, T. Reiss, C. Kehlet, T. Schulte-Herbr\"{u}ggen, and S. J. Glaser, Journal of Magnetic Resonance 172, 296 (2005).
\bibitem{PhysRevA.78.012328}C. A. Ryan, C. Negrevergne, M. Laforest, E. Knill, and R. Laflamme, Phys. Rev. A 78, 012328 (2008).

\bibitem{natcomm}A. M. Souza, J.-F. Zhang, C. A. Ryan, R. Laflamme, Nat. Commun. 2, 169 (2011).


\end{thebibliography}
\end{document}